\documentclass[twocolumn]{aastex63}

\usepackage{xspace}
\usepackage{comment}
\usepackage{float}

\def\starlight{\textsc{starlight}}                  

\newcommand{\msun}{\ifmmode \text{M}_{\odot} \else M$_{\odot}$\fi\xspace}

\newcommand{\rw}{ROGUE~I--WISE\xspace}

\newcommand{\Mstar}{\ifmmode M_{\star} \else $M_{\star}$\fi\xspace}
\newcommand{\MBH}{\ifmmode M_\text{BH} \else $M_\mathrm{BH}$\fi\xspace}
\newcommand{\Lrad}{\ifmmode L_{\rm 1.4} \else $L_{\rm 1.4}$\fi\xspace}
\newcommand{\Frad}{\ifmmode F_{\rm 1.4} \else $F_{\rm 1.4}$\fi\xspace}
\newcommand{\LHa}{$L_{\Ha}$\xspace}

\newcommand{\hii}{\ifmmode \rm{H}\,\textsc{ii} \else H~{\sc ii}\fi\xspace}
\newcommand{\Ha}{\ifmmode {\rm H}\alpha \else H$\alpha$\fi\xspace}
\newcommand{\Hb}{\ifmmode {\rm H}\beta \else H$\beta$\fi\xspace}
\newcommand{\oiii}{\ifmmode [\rm{O}\,\textsc{iii}] \else [O~{\sc iii}]\fi\xspace}
\newcommand{\oii}{\ifmmode [\rm{O}\,\textsc{ii}] \else [O~{\sc ii}]\fi\xspace}
\newcommand{\oi}{\ifmmode [\rm{O}\,\textsc{i}] \else [O~{\sc i}]\fi\xspace}
\newcommand{\nii}{\ifmmode [\rm{N}\,\textsc{ii}] \else [N~{\sc ii}]\fi\xspace}
\newcommand{\sii}{\ifmmode [\rm{S}\,\textsc{ii}] \else [S~{\sc ii}]\fi\xspace}
\newcommand{\Oiii}{[O~{\sc iii}]$\lambda$5007\xspace}
\newcommand{\Nii}{[N~{\sc ii}]$\lambda$6584\xspace}

\received{\today}
\revised{\today}
\accepted{\today}
\submitjournal{ApJ}

\shorttitle{Radio AGNs in the ROGUE I catalog using MIRAD}
\shortauthors{D. Kozie\l-Wierzbowska}

\begin{document}

\title{Identifying radio active galactic nuclei among radio-emitting galaxies}

\correspondingauthor{D. Kozie\l-Wierzbowska}
\email{dorota.koziel@uj.edu.pl}

\author[0000-0003-4323-0984]{D. Kozie\l-Wierzbowska}
\affiliation{Astronomical Observatory, Jagiellonian University, ul. Orla 171, PL-30244 Krakow, Poland}

\author[0000-0003-0842-8688]{N. Vale Asari}
\altaffiliation{Royal Society--Newton Advanced Fellowship}
\affiliation{Departamento de F\'{\i}sica--CFM, Universidade Federal de Santa Catarina, C.P.\ 476, 88040-900, Florian\'opolis, SC, Brazil}
\affiliation{School of Physics and Astronomy, University of St Andrews, North Haugh, St Andrews KY16 9SS, UK}

\author[0000-0002-4051-6146]{G. Stasi\'nska}
\affiliation{LUTH, Observatoire de Paris, PSL, CNRS 92190 Meudon, France}

\author[0000-0001-7907-7884]{F. R. Herpich}
\affiliation{Departamento de F\'{\i}sica--CFM, Universidade Federal de Santa Catarina, C.P.\ 476, 88040-900, Florian\'opolis, SC, Brazil}
\affiliation{Instituto de Astronomia, Geof\'{\i}sica e Ci\^{e}ncias Atmosf\'{e}ricas, Universidade de S\~{a}o Paulo, R. do Mat\~{a}o 1226, 05508-090 S\~{a}o Paulo, Brazil}

\author[0000-0003-1667-7334]{M. Sikora}
\affiliation{Nicolaus Copernicus Astronomical Center, Bartycka 18, 00-716 Warsaw, Poland}

\author[0000-0003-2644-6441]{N. \.Zywucka}
\affiliation{Centre for Space Research, North-West University, Potchefstroom 2520, South Africa}

\author[0000-0002-2224-6664]{A. Goyal}
\affiliation{Astronomical Observatory, Jagiellonian University, ul. Orla 171, PL-30244 Krakow, Poland}


\begin{abstract}

Basing our analysis on ROGUE I, a catalog of over 32,000 radio sources associated with optical galaxies, we provide two diagnostics to select the galaxies where the radio emission is due to an active galactic nucleus (AGN). Each of these diagnostics  can be applied independently. The first one, dubbed MIRAD,  compares the flux $F_{W3}$ in the $W3$ mid-infrared band of the WISE telescope,  with the radio flux at 1.4 GHz, $\Frad$. MIRAD requires no optical spectra. The second diagnostic, dubbed DLM, relates  the  4000 \AA\ break strength, $D_{\rm n}(4000)$, with the radio luminosity per unit stellar mass. The DLM diagram has already been used in the past, but not as standalone. 
For these two diagrams, we propose simple, empirical dividing lines that result in the same classification   for the objects in common. These lines correctly classify as radio-AGN 99.5 percent of the extended radio sources in the ROGUE~I catalog, and as star-forming (SF) galaxies 98--99 percent of the galaxies identified as such by their emission line ratios. 
Both diagrams clearly show that radio AGNs are preferentially found among elliptical galaxies and among galaxies hosting the most massive black holes.
Most of the radio sources classified  as radio-AGNs in the MIRAD or DLM diagrams are either optically weak AGNs or retired galaxies. 

\end{abstract}


\section{Introduction}
\label{intro}

One of the requirements for the study of nuclear activity in galaxies and in particular of the genesis and nature of radio jets that are present in only a minority of active galactic nuclei (AGNs) is the availability of large and complete multi-wavelength catalogs. In the last two decades several such catalogs have been produced by cross-matching optical, especially the Sloan Digital Sky Survey \citep[SDSS; ][]{York00}, and radio surveys, especially the First Images of Radio Sky at Twenty Centimetre \citep[FIRST; ][]{White97} and the NRAO  VLA  Sky  Survey \citep[NVSS; ][]{Condon98},  
\citep[e.g.][]{McMahon.etal.2002a,Sadler.etal.2002a,Best.etal.2005a,Kimball.Ivezic.2008a,Best.Heckman.2012a,Banfield.etal.2015a,Williams.etal.2019a,Sabater19,Koziel.Goyal.Zywucka.2020a}. 

The catalog of \citet{Koziel.Goyal.Zywucka.2020a} (ROGUE I, for `Radio sources associated with Optical Galaxies and having Unresolved or Extended morphologies I')
as it is published makes no claim about the nature of the radio emission, and contains radio emitting AGNs \citep[both jetted and non-jetted, according to the nomenclature promoted by ][]{Padovani17} and galaxies whose radio emission is related to recent star formation. While galaxies that clearly present 
resolved radio structures, such as jets and lobes, are known to inhabit an AGN, the question is more subtle (and difficult) for unresolved radio sources. 
A similar situation occurs with the on-going LOFAR Two-Metre Sky Survey (LoTSS) by \citet{Shimwell2017}, except that, in this case, optical spectra are not available for most sources. 
The aim of this paper is to present a simple way to distinguish radio sources associated with AGNs from those associated with star formation,  and to discuss the global properties of radio-emitting galaxies. 

Different groups in the past
 used a conjunction of several criteria to separate radio AGNs from star forming galaxies; e.g. \citet{Best.Heckman.2012a}, whose catalog matched SDSS DR\,7 galaxies with NVSS and FIRST radio sources using an automatic procedure and going down to 5 mJy, and \citet{Sabater19}, who base their sample on the first release of LOFAR data \citep{Shimwell.etal.2019} that go much deeper in radio. We prefer using just one criterion, which results into a  simpler classification and makes it easier to understand selection biases in the subsequent analyses. 
 
For samples of radio sources for which no optical spectra are available, we use the mid-infrared (MIR) data from the Wide-field Infrared Survey Explorer (WISE) survey \citep{Wright10} to distinguish radio emission linked with star formation from radio emission produced by active galactic nuclei. For samples of radio sources for which spectra of the host galaxies exist, we show that the diagram of the 4000 \AA\ break strength, $D_{\rm n}(4000)$, versus radio luminosity per unit stellar mass, $\Lrad/\Mstar$, herafter the DLM diagram, first introduced by \citet{Best2005} and revised several times \citep[see][]{Sabater19}, is on its own fully adequate to sieve out the radio-AGNs.

The paper is organized as follows. Section \ref{data} introduces the data used in this study. Section \ref{separating} presents the two alternative diagrams we propose to extract radio AGNs. Section \ref{other}  discusses the connection between the classifications presented in this paper and other galaxy classifications. Section \ref{summary} summarises our results.

\section{Data}
\label{data}

\subsection{Radio data}
\label{radiodata}

Our starting sample of radio emitting galaxies consists of the 32,616 objects presented in the ROGUE~I catalog. This catalog was made by selecting galaxies from the SDSS DR\,7 (see below) having a FIRST radio counterpart within 3 arcsec from the optical position. It provides information about core and total radio flux density at 1.4~GHz, as well as on the morphology of the radio structure and of the optical host galaxy. There are 1537 galaxies clearly connected to extended radio sources, 940 galaxies connected with a possibly extended, or one-sided structure, 902 sources considered as star-forming, blended, or not related to a galaxy with an SDSS spectrum. As many as  $29\,237$ objects fall into the unresolved or 'elongated' \citep[as opposed to 'extended', see][]{Koziel.Goyal.Zywucka.2020a} radio source categories.

The monochromatic radio luminosity  at 1.4~GHz is calculated from the total radio flux density ($\Frad$) as
  \begin{equation}
    \Lrad = \frac{4 \pi D_L^2}{(1+z)^{1 - \alpha}} \Frad,
  \end{equation}
  where we take the spectral index to be $\alpha = 0.75$ which is close to the mean spectral index of sources detected at 1.4 GHz \citep[e.g.][]{Condon1984a,Smolcic2017a}, and the luminosity distance $D_L$ is calculated from the spectroscopic redshifts from SDSS.  We assume a flat $\Lambda$CDM cosmology, with $\Omega_0 = 0.3$ and $H_0 = 70\,\text{km}\,\text{s}^{-1}\,\text{Mpc}^{-1}$.

\subsection{Optical data}
\label{opticaldata}

Objects in the ROGUE~I catalog have been selected from the SDSS seventh data release (DR7) database \citep{Abazajian.etal.2009a} and belong  
either to the Main Galaxy Sample \citep{Strauss.etal.2002a} or to the Luminous Red Galaxy sample \citep{Eisenstein01}. 
Note that, in principle, these samples do not contain any Type I AGNs (although \citealp{Oh2015a} identified around 4000 galaxies with broad \Ha\ lines in the full DR7 galaxy catalog).

The ROGUE~I catalog contains only SDSS DR7 galaxies with redshift $z$ larger than 0.002 (which guarantees that luminosity distances are not dominated by peculiar motions, \citealp[e.g.][]{Ekholm.etal.2001a}), and signal-to-noise ($S/N$) in the continuum at 4020 \AA\ of at least 10, in order to allow a meaningful study of their stellar populations. Repeated spectra of the same galaxy were removed by matching galaxies in right ascension and declination ($\Delta<0.5$ arcsec). From the ROGUE I catalog, we removed the small fraction of galaxies (0.2 per cent) for which the Petrosian half-light radius is negative or the stellar mass is smaller than $10^7 \,\mathrm{M}_\odot$ in the \starlight\ database, as well as the few objects that do appear in the published ROGUE I catalog but whose radio emission was found to be blended with another source or for which the optical galaxy is not the host of the radio emission (objects tagged B or ND in the catalog). Our final sample  contains 32163 galaxies, and is referred to as the ROGUE~I sample in the remaining of the paper.

The SDSS spectra allow us to derive certain physical quantities such as the redshifts, the total galaxy stellar masses (\Mstar), the black hole masses (\MBH), and $D_{\rm n}(4000)$.  These last quantities are obtained by means of the inverse spectral synthesis code \starlight\ \citep{CidFernandes.etal.2005a} to fit the observed continuum emission of the galaxies.
The black hole masses are derived from the stellar velocity dispersions ($\sigma_\star$) using the relation by \citet{Tremaine.etal.2002a}, which is applicable for $\sigma_\star \geq 70$ km s$^{-1}$. $D_{\rm n}(4000)$ is taken from the synthetic spectra  as explained in \citet{Stasinska.etal.2006a}.
Emission-line measurements were performed after subtraction of the synthesized stellar spectrum from the observed SDSS spectrum. All the data were obtained exactly as explained in section~2 of \cite{KozielWierzbowska.etal.2017a}, and can be retrieved from the \starlight\ database\footnote{http://www.starlight.ufsc.br} \citep{Fernandes09}.  
 
\subsection{Mid-infrared data}
\label{mirdata}

WISE observed the entire sky in four infrared bands with central wavelengths 3.4, 4.6, 12 and 22 $\mu$m \citep[dubbed $W1$, $W2$, $W3$, and $W4$, respectively;][]{Wright10}. The photometric sensitivity at $5 \sigma$ is 0.068, 0.098, 0.86 and 5.4 mJy for each one of the four bands. The catalog contains positional data of more than 500 million objects with $S/N > 5$ for at least one of the bands. The photometric calibration was made in the Vega system, with conversion factors to the AB system of 2.699, 3.339, 5.174 and 6.620 for $W1$, $W2$, $W3$, and $W4$, respectively \citep{Jarrett11}.

Emission in the $W1$ and $W2$ bands is more easily detected because of the sensitivity of the detector and the data usually have better signal-to-noise. The widest band is $W3$, with a spectral coverage of 7--17$\mu$m, which was designed to observe the strong Polycyclic Aromatic Hydrocarbon (PAH) emission bands that are present in this range \citep{Jarrett11}. The PAH emission is often associated with the presence of warm dust within molecular clouds, excited by the strong radiation field from the massive young stars immersed in it \citep{daCunha.etal.2008a} and/or from the dust torus around an AGN \citep{Alonso-Herrero.etal.2014}. The main features of these kinds of emission are centred around $11\mu$m, hence detectable inside the $W3$ band for the whole redshift range probed by the ROGUE I sources ($z \leqslant 0.6$), while thermal emission is found within the $W4$ band, the most $S/N$-poor filter of WISE \citep{Wright10}.

We matched the WISE and the SDSS DR\,7 catalogs, identifying the most probable pairs within 1 arcsec radius. 
To use the WISE measurements, we required $S/N_{W3} \geqslant 2$ to exclude spurious measurements and problems due to sensitivity. This is especially needed for early-type galaxies, which have little or no gas and are generally very faint in this spectral range. 
 
Our final sample of galaxies with usable WISE data consists of $23\,294$ objects, and  will be referred to as the \rw sample.
 
Throughout the text we also use the flux calculated for the $W3$ band.  We obtain fluxes from Vega magnitudes using \begin{equation}
    F_\nu \,\mathrm{[Jy]} = F_{\nu 0} 10^{-0.4 m_\mathrm{Vega}},
\end{equation}
where $F_{\nu 0}$ is the zero magnitude flux density and $m_\mathrm{Vega}$ is the WISE magnitude of the object. 
We use the zero points and the relations from \citet{Jarrett11}\footnote{And also from the WISE website, \url{http://wise2.ipac.caltech.edu/docs/release/allsky/expsup/sec4_4h.html}.} with a zero magnitude flux density $F_{\nu 0} = 31.674$ for $W3$ and assuming that the MIR spectral energy distributions do not significantly deviate from a power-law ($F_\nu \sim \nu^0$).


\section{Separating radio AGNs from star-forming galaxies}
\label{separating}

\subsection{The advantage of using only one criterion}
\label{onecri}

As mentioned in the introduction, we look for a unique criterion which would be able to separate star-forming (SF) galaxies from galaxies where the radio emission is linked to an AGN. The first problem with using several criteria together, as is the case of e.g.\ \citet{Best.Heckman.2012a}, is that relevant data are not always available for all the objects, so that for some objects the classification has to rely on only a fraction of the chosen criteria.  

We have matched their catalog of 18\,268 objects to the \starlight\ database, and found 18\,074 objects in common. To analyse their data we used the radio fluxes from the \citet{Best.Heckman.2012a} catalog, and emission lines and stellar masses and other spectral measurements from the  \starlight\ database.
Objects which do not exhibit the four emission lines (\oiii, \Hb, \nii\ and \Ha) in the \citet[][BPT]{Baldwin.etal.1981a} diagram cannot be classified using this criterion. This concerns 11\,939 (66 per cent) of their entire sample of $\sim 18$k objects.

Second, for those objects with data allowing one to apply more than one criterion, classifications according to the various criteria do not necessarily agree. 
Let us consider the other two criteria used by \citet{Best.Heckman.2012a}, which are $D_{\rm n}(4000)$ versus $\Lrad/\Mstar$, and the relation between the \Ha\ emission-line luminosity and the radio luminosity ($L_{\Ha}$ versus \Lrad).  Among the 9\,596 objects from \citet{Best.Heckman.2012a} having all the required data to use the two methods, there are 1\,085 objects for which the classifications differ. 

In the following subsections we present two methods which can be individually applied to radio catalogs depending on data availability.

\subsection{The MIRAD diagram}
\label{mirad}

It has been known almost since the beginning of extragalactic radio astronomy and infrared astronomy  \citep[][and references therein]{deJong.etal.1985a,Helou.etal.1985a,Condon.Broderick.1988a,Condon.1992a} and confirmed in recent studies \citep[e.g.][]{Wang2019a}  that there is a tight correlation between the radio luminosity and the far infrared luminosity of `normal' star forming galaxies, i.e.\ galaxies that do not contain an AGN. The radio emission in these sources consists of free-free radiation from \hii\ regions and synchrotron radiation from supernova remnants (mostly Type II and Type Ib) whose progenitors are more massive than 8 \msun, so it probes stellar populations younger than $5\times 10^7$--$10^8$ yr. The far infrared radiation measures the bolometric luminosity of stars more massive than about 5 \msun  reprocessed by circumstellar dust. Neither the radio emission nor the far infrared emission are affected by dust extinction, thus the observed radio flux and far infrared flux are expected to be tightly linked. In the local Universe (up to $z = 0.15$), the relation is very tight, with a dispersion of about 0.25 dex \citep{Yun2001a}. It has been used to identify jetted AGNs as objects deviating from this relation \citep{Donley2005a,Park2008a,DelMoro2013a}. However, existing far-infrared data are either from rather shallow surveys (IRAS, \citealp{Neugebauer1984a}, or Infrared Space Observatory, ISO) not deep enough to be useful for the characterization of ROGUE I galaxies, while Spitzer \citep{Werner2004a} data are available only for selected objects.

\begin{figure} 
  \centering
  \includegraphics[width=0.85\columnwidth, trim=10 15 10 10, clip]{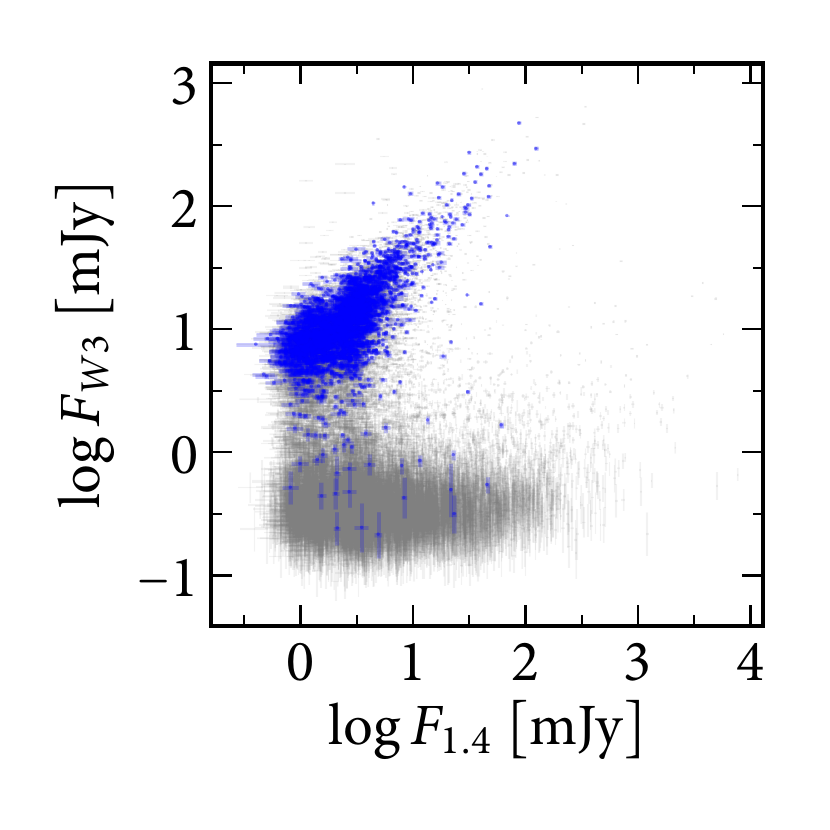}
  \caption{$\log F_{W3}$ versus $\log \Frad$ (i.e. the  MIRAD diagram) for the  objects in the \rw sample (see text). The full sample is shown as grey crosses, whose sizes are the uncertainties in log fluxes. Points in blue are the $2\,549$ pure SF galaxies as defined by  \citet{Stasinska.etal.2006a} on the \Nii/\Ha versus \Oiii/\Hb plane.
  }
  \label{fig:mirad-sf}
\end{figure}

On the other hand, mid-infrared surveys exist for the whole sky (AKARI, \citealp{Murakami2007a}, and WISE, \citealp{Wright10}). \citet{Rosario13} showed the luminosity in the $W3$ band versus \Lrad for a sample of emission-line AGNs selected from the SDSS galaxy data base. They found that these emission-line AGNs were distributed into two branches. Seyfert galaxies gather on the upper branch, in the same region as a control sample of inactive star-forming galaxies. Low-ionization nuclear emission line galaxies (LINERs) are split between the upper branch and the lower branch. These authors argued that, for the Seyfert galaxies, the $W3$ emission is almost entirely due to dust heated by ongoing star formation and highlighted the MIR--radio plane as a useful tool in studies of star formation and accretion properties of AGNs. The MIR--radio plane was also successfully used to study star-formation properties of the radio/X-ray selected sample by \citet{Mingo2016}.

Fig.~\ref{fig:mirad-sf} shows the ROGUE I--WISE sample on the $\log F_{W3}$ versus $\log \Frad$ plane, which we dub MIRAD as it is based on MId-infrared and RADio fluxes. The full sample is plotted as grey error bars, which have been obtained by error propagation from the observed fluxes. Points in blue highlight the $2\,538$ objects which are classified as pure star-forming galaxies according to \citet{Stasinska.etal.2006a} on the \Nii/\Ha versus \Oiii/\Hb plane (requiring a minimum $S/N$ of 3 in those emission lines). Several features are noteworthy. First, this plot shows two different families of objects, with SF galaxies falling on the upper sequence. Second, uncertainties are much smaller in $F_{W3}$ than in \Frad for the SF sequence, and the opposite is true for the lower sequence.

This diagram shows a very clear dichotomy among the population of galaxies with radio emission. We looked for objective criteria to draw a separation line. 
It turns out that an attempt to separate the BPT-pure SF galaxies and the classical extended radio galaxies \citep[FR~I and FR~II,][]{Fanaroff.Riley1974a} gives an infinity of solutions which are equally valid. We thus chose the simplest of all, namely 
\begin{equation}
  F_{W3} =\Frad ,
\end{equation}
where both quantities are in mJy. 

\begin{figure} 
  \centering
  \includegraphics[width=0.9\columnwidth, trim=10 15 10 20, clip]{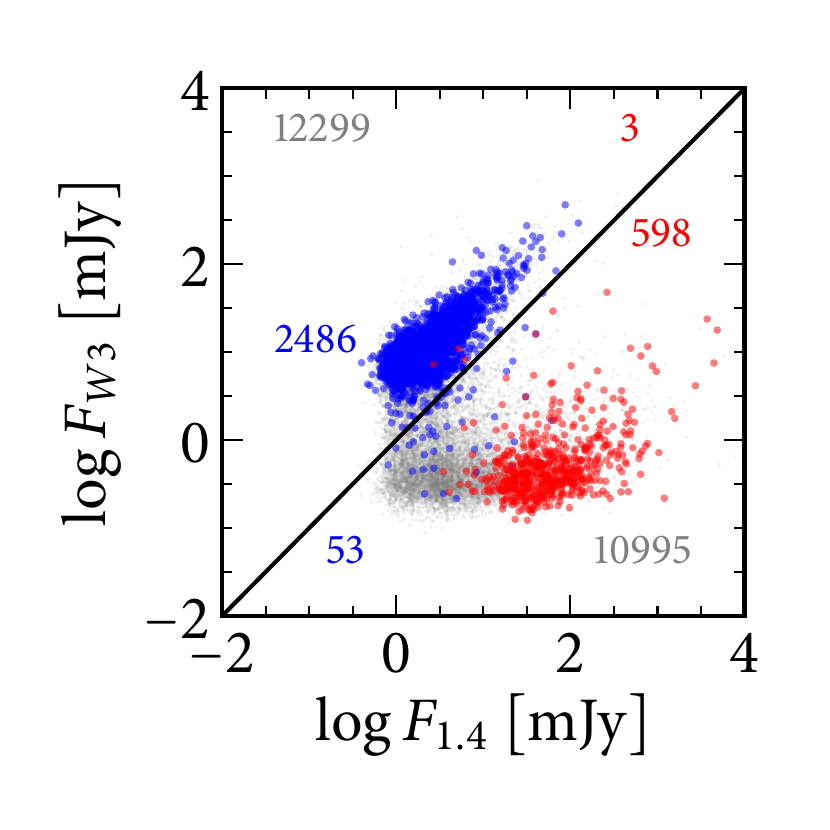}
\caption{$\log F_{W3}$ versus $\log \Frad$. The $23\,294$ objects from the \rw sample are in grey. In blue we show the BPT-pure SF, and in red the FR~I and FR~II radio sources. The continuous line represents the limit below which we consider that all the galaxies are radio-AGNs. Numbers in grey, red and blue show how many of the ROGUE I--WISE, FR~I and FR~II, and BPT-pure SF  objects fall above and below the dividing line.  
}
  \label{fig:mirad}
\end{figure}

This line is drawn  in Fig.~\ref{fig:mirad}, where the pure SF galaxies are plotted as blue points and the objects corresponding the classical extended radio galaxies (FR~I and FR~II) are plotted as red points. All but 3 objects out of a total of 601 from the FR~I, FR~II class  lie below this line. In the remaining of the paper, we consider that all the galaxies below this line are radio AGNs, meaning that their radio emission is dominated by phenomena related to an AGN (mainly a jet originating from a supermassive black hole), and that only a small fraction of it, if any, is due to star-forming regions.
 The three FR~II radio galaxies that are above the dividing line are most probably misclassifications. We have checked them and, while their NVSS structures are considered as lobes in ROGUE I, they may be in fact separate sources. 
Objects above the line are considered to be galaxies where the radio emission is mainly due to star formation.
Note that there are however 53 objects out of 2539 (2 per cent) that are classified as SF according to the \citet{Stasinska.etal.2006a} line in the BPT diagram but lie below the separation line, some of them even quite far from it. They may actually not be \textit{pure\textit{}} SF galaxies, since one has to recall that the SF demarcation line is actually an \textit{upper} envelope for ionization by massive stars.  Possibly these objets do contain an active AGN. Detailed studies of them should provide more information about their nature.

According to our MIRAD dividing line, $10995/(10995 + 12229)$, i.e. roughly 47 per cent of the objects from the \rw sample are radio AGNs. 
Note that the MIRAD discriminator can be applied without any knowledge of the redshift, which can be useful for surveys such as LOFAR that do not involve optical spectra.

\subsection{The $D_{\rm n}(4000)$ versus $\Lrad/\Mstar$ (DLM) diagram}
\label{dlm}

\begin{figure} 
  \includegraphics[width=0.9\columnwidth,, trim=0 0 0 0, clip]{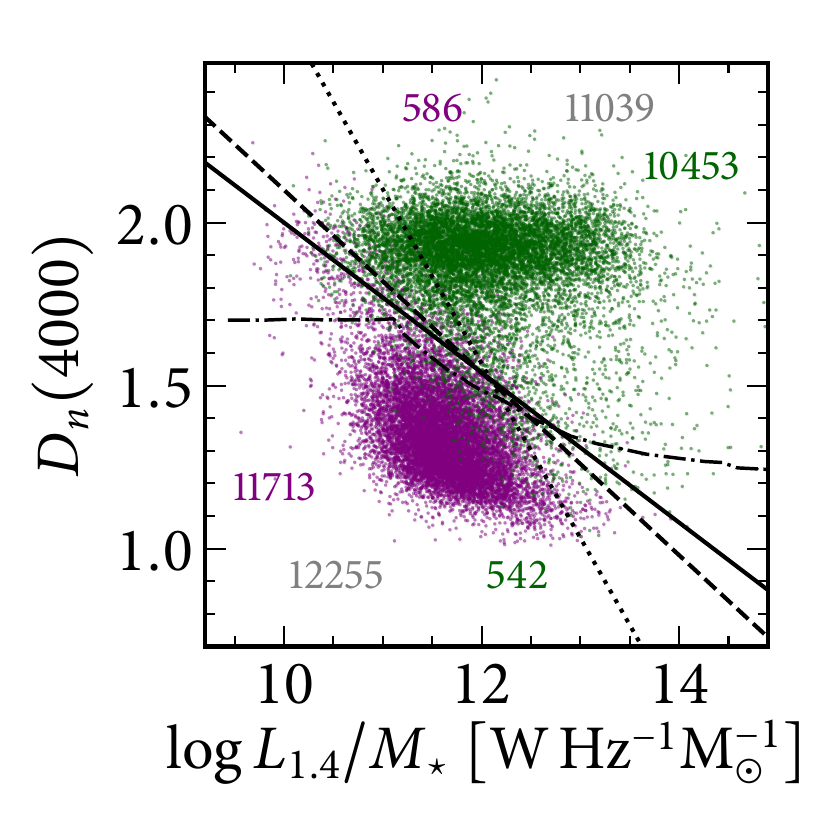}
  \caption{The $D_{\rm n}(4000)$ versus $\Lrad/\Mstar$ (DLM) diagram for the \rw sample. 
    Purple points: SF galaxies according to the MIRAD diagram, green points: radio-AGNs according to the MIRAD diagram. Thin lines: some previous dividing lines (dots: \citealp{Kauffmann+2008}, dot-dashes: \citealp{Sabater19}).  Thick lines: dividing lines that minimize the difference with the MIRAD diagnostic (continuous: our first choice, Eq. \ref{eq:dlm}; dashes: our second choice). In the following, all the diagnostics in the DLM diagram will be based on the continuous line. The numbers indicate the total score of objects above and below the continuous dividing line; the total is in grey, MIRAD-SF in purple, and MIRAD-AGN in green.
  }
  \label{DLMrw}
\end{figure}

\begin{figure} 
  \includegraphics[width=0.9\columnwidth,, trim=0 0 0 0, clip]{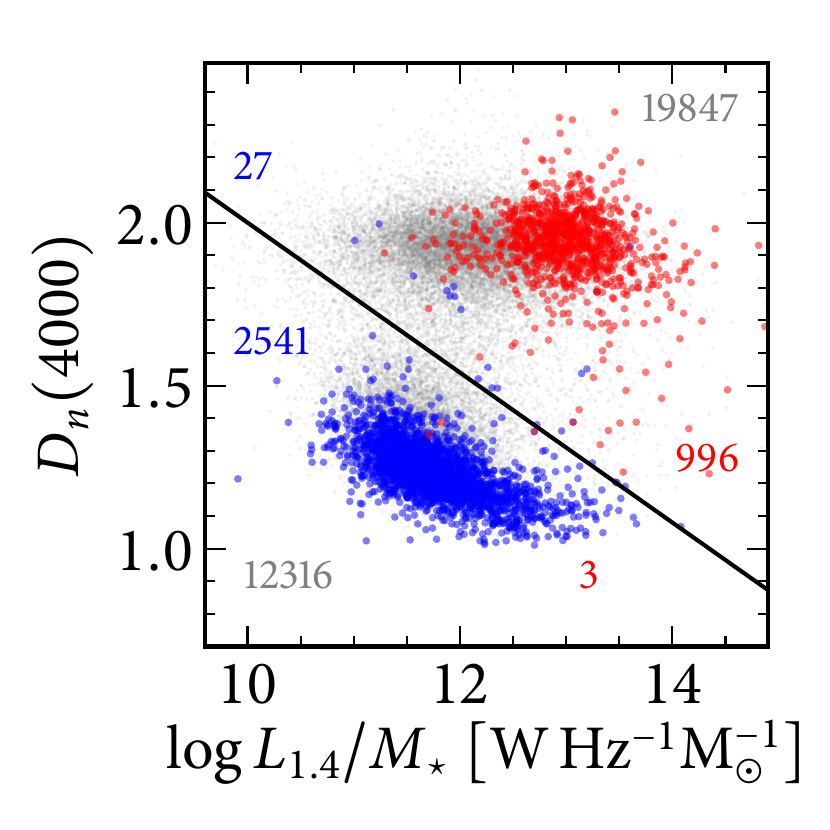}
  \caption{The DLM diagram for the entire ROGUE~I sample. Blue points: SF galaxies according to the BPT diagram. Red points: classical extended radio AGNs (FR~I and FR~II radio sources). }
  \label{DLMrogue1}
\end{figure}

Among the diagrams used by previous authors to select radio AGNs, the DLM one is quite efficient one and can be applied to objects with or without emission lines,  whether or not they have any WISE detections. Different dividing lines have been proposed in the past \citep{Best2005,Kauffmann+2008,Best.Heckman.2012a,Sabater19}. Here, since we emphasized the utility of a unique diagram, and since we have shown that the MIRAD diagram works extremely well when no spectra are available and no redshift can be determined, we defined a line for the DLM diagram that is as much compatible as possible with our dividing line in the MIRAD diagram. We proceeded as follows. 
  For a given dividing line in the DLM digram, we defined the completeness $C$ and the $R$ reliability fractions for the SF and AGN classes \citep{Strateva.etal.2001a}. $C_\mathrm{SF}$ is the fraction of galaxies classified as MIRAD-SF which are correctly classified as SF in the DLM diagram, whereas $R_\mathrm{SF}$ is fraction of galaxies classified as SF in the DLM diagram which are also MIRAD-SF.
  Defining  $C_\mathrm{AGN}$ and $R_\mathrm{AGN}$ likewise for the radio AGN class, we calculated the figure-of-merit $P$ \citep{Baldry.etal.2004a}, which is the product of those four fractions: $P \equiv C_\mathrm{SF} R_\mathrm{SF} C_\mathrm{AGN} R_\mathrm{AGN}$.
  We calculated $P$ over a wide range of slopes and intersepts for the DLM dividing line, refining our search in the loci of the greatest values of $P$.

Two possible solutions were found: 
\begin{equation}
D_{\rm n}(4000)  = -0.28\, \Lrad/\Mstar + 4.9,
\end{equation}
and
\begin{equation}
D_{\rm n}(4000)  = -0.23\ \Lrad/\Mstar + 4.3.
\label{eq:dlm}
\end{equation}
We chose equation \ref{eq:dlm}, which yields a marginally greater value of $P$, for the dividing line in the DLM diagram.

Fig.~\ref{DLMrw} shows the values of $D_{\rm n}(4000)$ as a function of  $\Lrad/\Mstar$ for the \rw sample, together with the dividing lines from \citet{Kauffmann+2008}, \citet{Sabater19}\footnote{Actually \citet{Sabater19} used the LOFAR radio luminosities at 150 MHz, which are related to the 1.4 GHz luminosities by $\log \Lrad \approx \log L_\mathrm{150 \, MHz} - 0.68$.} and our own dividing lines. Fig.~\ref{DLMrogue1} shows the DLM diagram for the ROGUE~I sample, together with the dividing line represented by Eq. \ref{eq:dlm}. Here, the red points correspond to the extended radio AGNs (FR~I and FR~II)  and the blue points to SF galaxies from the BPT diagram, as in Fig.  \ref{fig:mirad}. We see that our DLM dividing line puts only 3 extended radio AGNs (out of a total of 999) in the SF zone, and only 27 BPT SF galaxies out of a total of 2568 (i.e. one per cent) in the radio AGN zone. So the DLM diagram with our dividing line seems work even slightly better than the MIRAD diagram.

The total number of objects that can be classified with the DLM diagram is 32\,163 (as opposed to 23\,294 in the MIRAD diagram. However the ratio of the number of radio-AGN and SF galaxies is 1.61 in the DLM diagram for the ROGUE I sample and 0.89 in the MIRAD diagram for the \rw sample. In other words, many radio-AGNs are missing in the MIRAD diagram. Those are likely the ones that have too small flux in the $W3$ band to be detected. This means that the MIRAD diagram provides a biased estimate of the total number of radio-AGNs among radio sources. This needs to be remembered when dealing with demography of radio AGNs. A way to circumvent this would be to use upper limits for the $W3$ detection in order to cull a sample of objects with low MIR emission.


\section{MIRAD and DLM versus other classifications of galaxies}
\label{other}

\begin{figure*} 
  \centering
  \includegraphics[width=\textwidth, trim=20 0 50 0, clip]{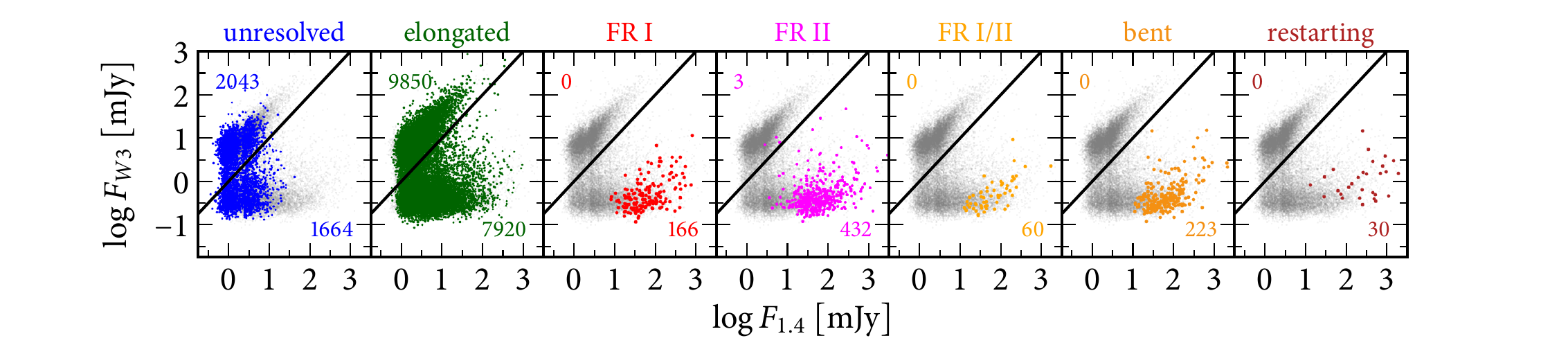}
  \caption{The distribution of the objects with different radio morphologies in the MIRAD diagram. The \rw sample is in grey. Sources with specific morphological types are marked in colours: unresolved sources in blue, elongated in green, FR~I in red, FR~II in magenta, hybrid in light yellow, bent sources (including wide-angle tail, narrow-angle tail and head-tail radio sources) in dark yellow, and restarting sources (which are double-double, X-shaped and Z-shaped sources) in brown. }
  \label{fig:miradradmorph}
  \includegraphics[width=1.09\textwidth, trim=15 0 0 0, clip]{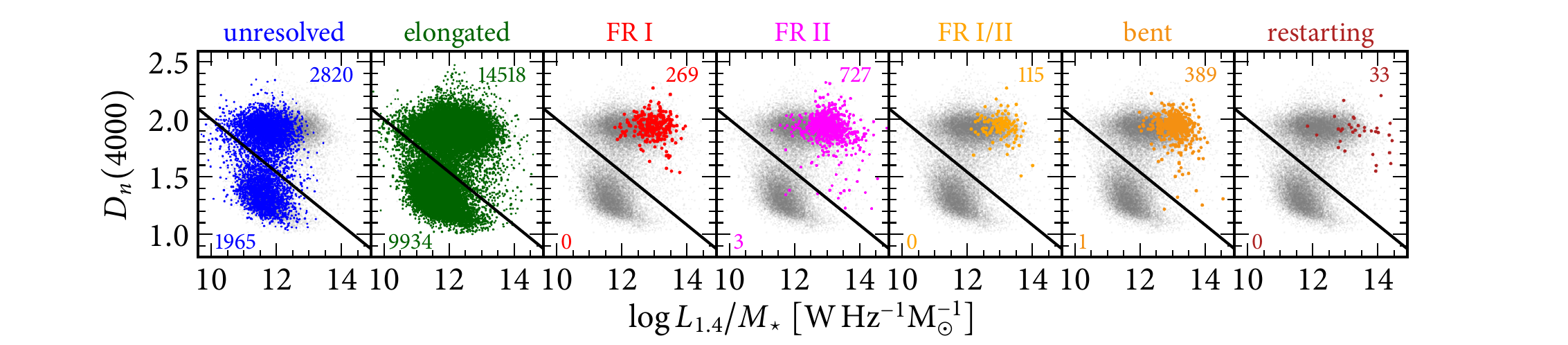}
  \caption{The same as Fig.~\ref{fig:miradradmorph} but for the DLM diagram and the ROGUE I sample.}
  \label{fig:dlmradmorph}
  \includegraphics[width=\textwidth, trim=20 0 50 0, clip]{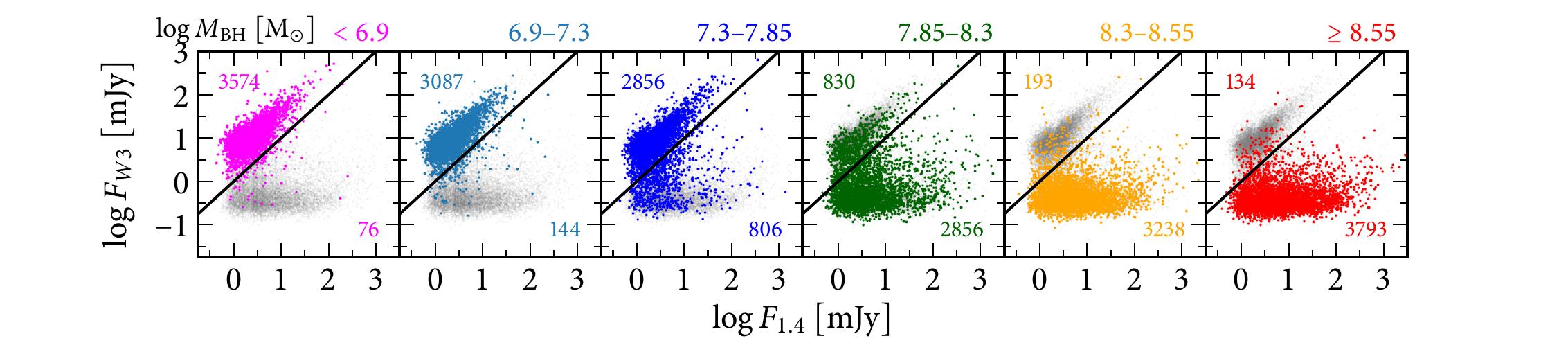}
  \caption{The MIRAD diagram in bins of BH mass. All the objects of the ROGUE I--WISE sample are represented as grey points. Superimposed in different colours are represented those object having a  BH mass within the limits indicated at the top of the panels. }
  \label{fig:miradBH}  
  \includegraphics[width=1.09\textwidth, trim=15 0 0 0, clip]{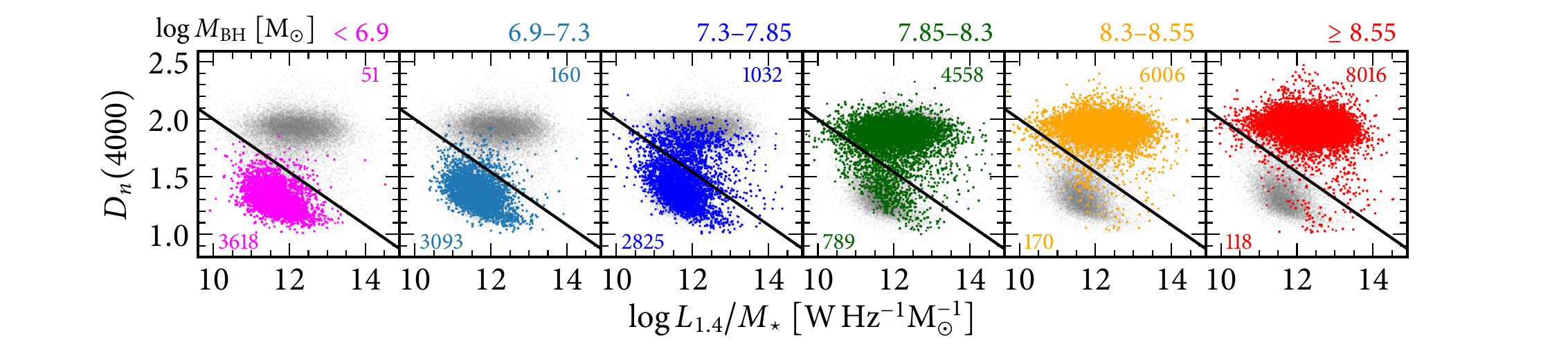}
  \caption{The same as Fig.~\ref{fig:miradBH} but for the DLM diagram and the ROGUE I sample.}
  \label{fig:DLMMBHrogue1}
\end{figure*}


\subsection{Radio classes}
\label{mirad-radio}

Fig.~\ref{fig:miradradmorph} shows the distribution in the MIRAD diagram of the radio morphologies as listed in the ROGUE I catalog. As expected, almost all the extended radio morphologies (i.e.\ FR~I, FR~II, hybrid, bent sources which are wide-angle tail, narrow-angle tail, and head-tail, and restarted radio sources comprised of double-double, X-shaped, and Z-shaped sources) are radio AGNs. 
Unresolved and elongated radio sources, on the other hand, are found on both sides of the dividing line, in similar proportions. It is precisely for those cases that our diagrams are useful to pinpoint which objects correspond to radio AGNs.

Fig.~\ref{fig:dlmradmorph} is the equivalent of Fig.~\ref{fig:miradradmorph} but for the DLM diagram applied to the ROGUE 1 sample. Qualitatively, the results are similar to those obtained with the MIRAD diagram. 

\subsection{Black hole masses for radio-AGNs}
\label{mirad-BH}

\begin{figure} 
  \centering
  \includegraphics[width=0.9\columnwidth, trim=40 0 390 0, clip]{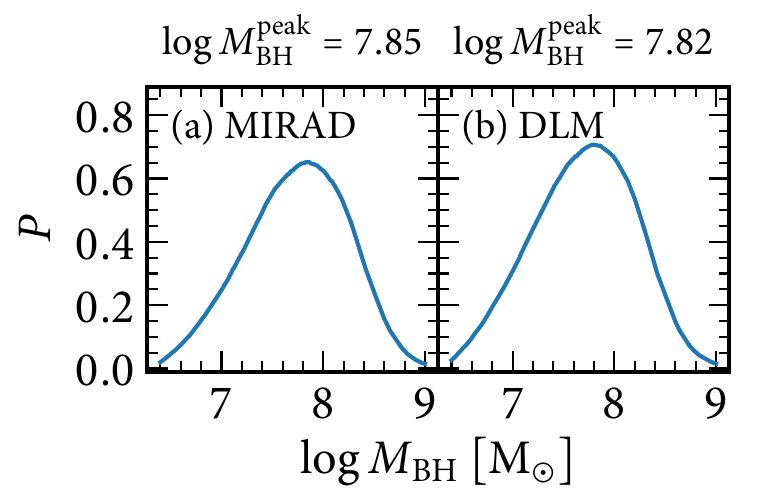}
  \caption{(a) The figure of merit $P$ as a function of \MBH, using the MIRAD diagram and the \rw sample. 
(b) The figure of merit $P$ as a function of \MBH, using the DLM diagram and the ROGUE~I sample. The peak indicates the optimal \MBH value that separates radio emission from SF or an  AGN.}
  \label{mirad_MBH_lim}
\end{figure}

It has already been shown in many studies \citep[e.g.][]{Dunlop2003a,KozielWierzbowska.etal.2017a} that the fraction of radio-loud AGNs increases rapidly in galaxies with black holes more massive than $\log$\,\MBH/$\msun \sim 8$. Fig.~\ref{fig:miradBH} shows the distribution of \rw galaxies with black hole masses in different ranges in the MIRAD diagram, and Fig.~\ref{fig:DLMMBHrogue1} shows the same for ROGUE~I in the DLM diagram.  1\,727 objects in the ROGUE~I and 1\,707 in the \rw samples have $\sigma_\star < 70$ km~s$^{-1}$, so they lack \MBH estimates and are not included in any \MBH bin.

We note that galaxies with low \MBH fall preferentially  above the MIRAD line, and those with high \MBH fall below it, suggesting there is an optimal \MBH separation between SF and radio AGN. To find which \MBH value best corresponds to our MIRAD separation line, we again use the the figure-of-merit $P$ (see Section \ref{dlm}), where the completeness $C$ and the $R$ reliability fractions are defined according to classification as SF or AGN by MIRAD and a \MBH limit. Fig.~\ref{mirad_MBH_lim} shows $P$ as a function of the \MBH threshold. For the MIRAD diagram and the \rw sample (panel a), $P$ reaches its maximum value at $\log$\,\MBH/$\msun = 7.85$. For the DLM diagram applied to the ROGUE I sample (panel b), $P$ reaches its maximum value at $\log$\,\MBH/$\msun = 7.82$.

Thus, both the MIRAD and the DLM diagrams indicate that radio AGNs are found almost exclusively among galaxies with black hole masses above $10^{7.8}$ \msun.


\subsection{Galaxy morphological classes}
\label{mirad-optmorph}

\begin{figure*} 
  \centering
  \includegraphics[width=\textwidth, trim=20 150 20 0, clip]{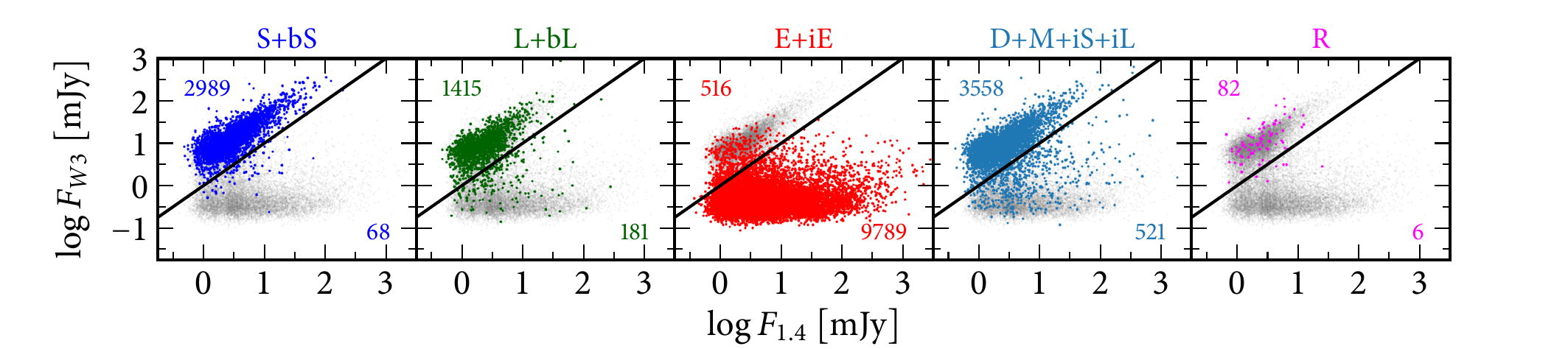}
  \caption{The distribution of the objects with different optical morphologies in the MIRAD diagram. The full \rw sample is in grey. Different morphological types are shown in different colours: spiral and barred spiral in blue, lenticular and barred lenticular in green, elliptical and interacting elliptical in red, distorted, merger, and interacting spiral and lenticulars in light navy, and ring in magenta. }
  \label{fig:miradoptmorph}
\end{figure*}

\begin{figure*} 
  \includegraphics[width=1.02\textwidth, trim=15 150 0 0, clip]{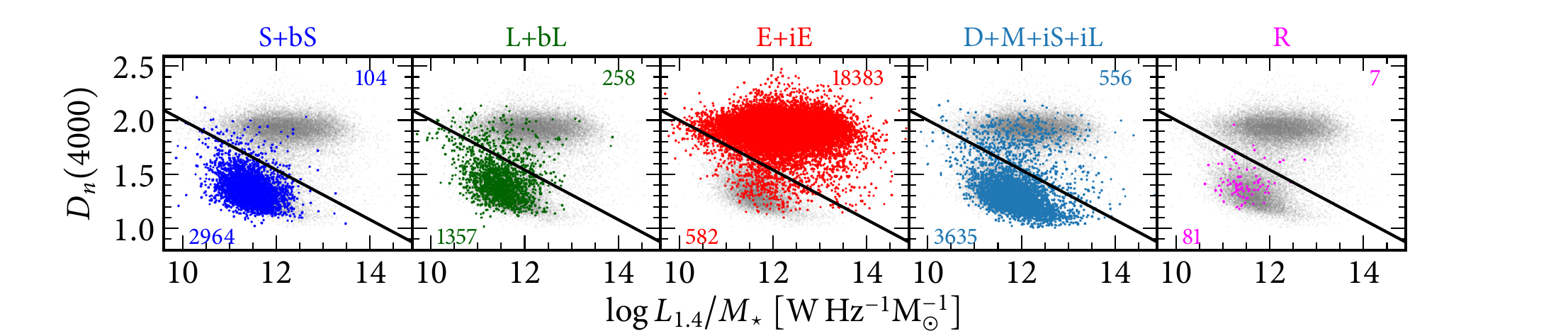}
  \caption{The same as Fig. \ref{fig:miradoptmorph} but for the DLM diagram and the ROGUE I sample.}
  \label{DLMOptmorph}
\end{figure*}

Fig.~\ref{fig:miradoptmorph} shows the distribution in the MIRAD diagram of the optical morphological classifications from the ROGUE I catalog. Spiral (S) and barred spiral (bS) galaxies are shown in blue, lenticular (L) and barred lenticular (bL) galaxies are shown in green, elliptical (E) and interacting elliptical (iE) in red, galaxies with distorted morphology (D), mergers (M), interacting spiral (iS), and interacting lenticular (iL) galaxies are shown in light navy blue, while ring (R) galaxies are marked in purple. 

Later-type galaxies (spirals and lenticulars) fall above the MIRAD separation line, whereas earlier-type ones (ellipticals) fall below the line.
Galaxies showing signs of interaction also lie above the separation line, which suggests that their radio emission is connected to star formation.
About 13 per cent of interacting systems however are below the line, which is in agreement with the fact that a significant fraction of radio jets are found in mergers and interacting galaxies \citep[see e.g.][]{Chiaberge2015a}.

Fig.~\ref{DLMOptmorph} is similar to Fig.~\ref{fig:miradoptmorph} but for the DLM criterion applied to the ROGUE I sample. Similarly to the MIRAD diagram, and even more conspicuously, it shows that the vast majority of radio-AGNs are found among elliptical galaxies.


\subsection{Galaxy spectral classes}
\label{mirad-spectral}

\begin{figure*}
  \centering
  \includegraphics[width=\textwidth, trim=20 0 20 0, clip]{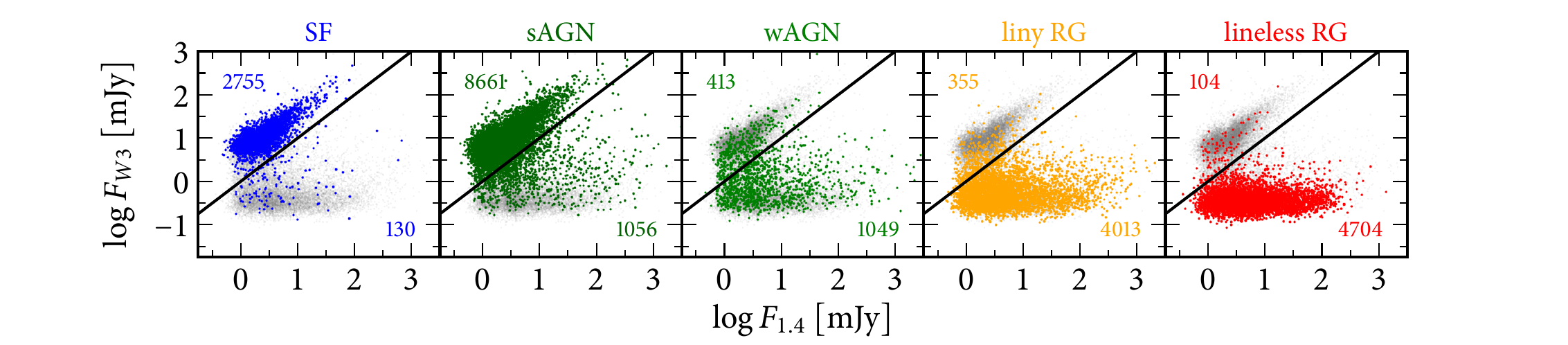}
  \caption{The distribution of the objects with the different WHAN spectral classifications in the MIRAD diagram. The full \rw sample is in grey. The WHAN categories are represented in different colours: pure SF galaxies in blue, strong AGNs in dark green, weak AGNs in light green, liny RGs in orange, lineless RGs in red.}   
  \label{fig:miradwhan}
\end{figure*}

\begin{figure*} 
  \includegraphics[width=1.02\textwidth, trim=15 0 0 0, clip]{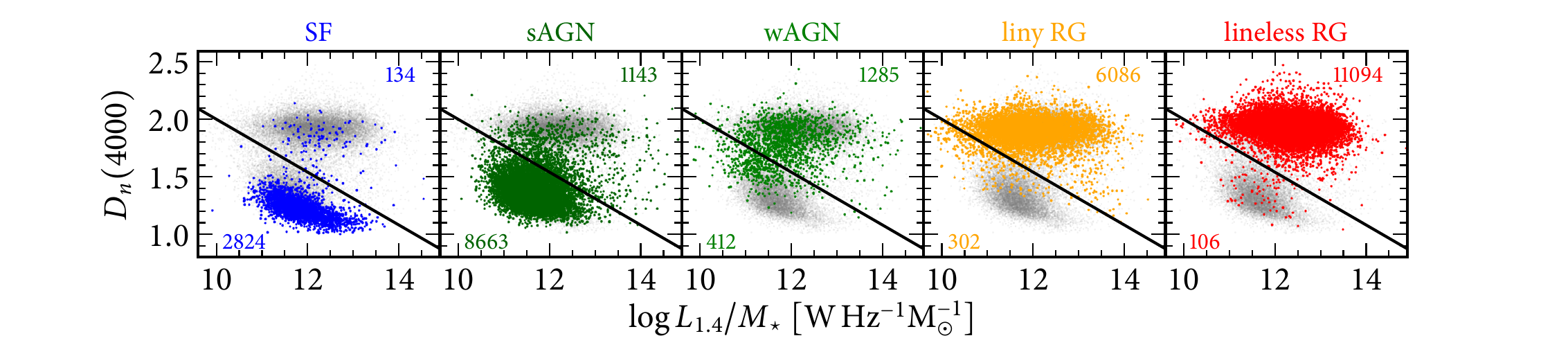}
  \caption{The same as Fig. \ref{fig:miradwhan} but for the DLM diagram and the ROGUE I sample.}
  \label{dlmwhan}
\end{figure*}

\begin{figure*} [h]
  \centering
  \includegraphics[width=0.9\textwidth, trim=10 10 10 10, clip]{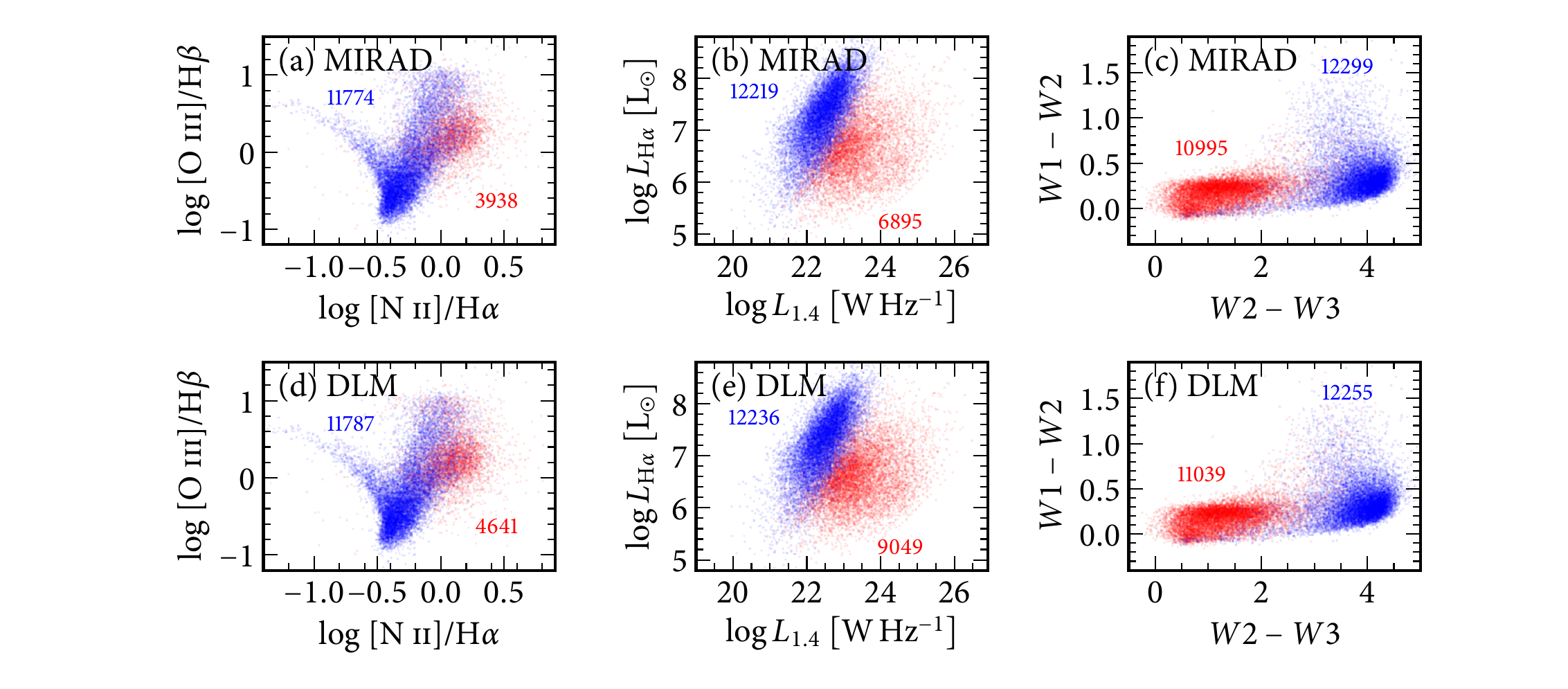}
  \caption{The distribution of objects in three complementary diagrams used in previous studies. \textbf{Top row:} colours according to the MIRAD classification of the \rw sample. \textbf{Bottom row:} colours according to the DLM classification of the ROGUE~I sample.  Radio AGNs are in red, the remaining objects are in blue.  The numbers in red and blue show the number of blue and red points in each diagram.} 
  \label{fig:classif}
\end{figure*}

Fig.~\ref{fig:miradwhan} shows the distribution of the five categories of galaxies defined by \citet{Fernandes2011} in the $W_{\Ha}$ versus \nii/\Ha\ diagram (the WHAN diagram): 
\begin{itemize}
  \item Pure star-forming galaxies: $\log\, \nii/\Ha\ < -0.4$ and  $W_{\Ha} \geq 3$ \AA, shown in blue,
  \item Strong AGNs (i.e.\ Seyferts): $\log\, \nii/\Ha \geq -0.4$ and  $W_{\Ha} \geq 6$ \AA, shown in dark green,
  \item Weak AGNs: $\log\, \nii/\Ha\ \geq -0.4$ and  $3 \leq W_{\Ha} < 6$ \AA, shown in light green,
  \item Liny retired galaxies (RGs, i.e.\ fake AGN): $0.5 \leq W_{\Ha} < 3$ \AA, shown in orange,
  \item Lineless RGs: $W_{\Ha} < 0.5$ \AA, shown in red. 
\end{itemize}

Note that the value of $\log\, \nii/\Ha\ = -0.4$, which separates pure SF-galaxies from AGNs in the WHAN classification of \citet{Fernandes2011} is based on the line proposed by \citet{Stasinska.etal.2006a} to separate \textit{pure} SF galaxies from AGNs in the BPT diagram. Objects with $\log\, \nii/\Ha\ > -0.4$ may well experience present-day star formation but the presence of an AGN can be detected in their optical spectrum (as compared to \textit{pure} SF galaxies), be it at a level of a few percent. 
Fig.~\ref{dlmwhan} is the same as Fig.~\ref{fig:miradwhan} but for the DLM diagram applied to the ROGUE~I sample. 

A few objects in our samples (54 in the \rw sample and  114 in the ROGUE~I sample) cannot be classified with the criteria above: They have $W_{\Ha} \geq 3$ \AA\ (i.e.\ they are not retired), but lack an $\nii/\Ha$ measurement, meaning they cannot be classified as either being an optical AGN host or a SF galaxy.

Figs.~\ref{fig:miradwhan} and \ref{dlmwhan} reveal that the SF and sAGN classes are distributed very similarly in the MIRAD and in the DLM diagram.  Only a small fraction of sAGNs are identified as radio-AGNs. The proportion is somewhat larger for wAGNs (which can broadly be identified with `true LINERs', i.e.\ galaxies with low-level nuclear activity). In both the MIRAD and the DLM diagrams, one can see that liny and lineless retired galaxies consitute the vast majority of radio-AGNs: 79 per cent in MIRAD and 87 per cent for the DLM diagram.
These percentages must not be taken literally of course, but they show that any classification based on emission lines alone would miss the huge population of radio-AGN which are lineless RGs.
 
Radio AGNs are often divided into high- and low-excitation radio galaxies (HERGs and LERGs, respectively), e.g.\ \citet{Laing1994a, Buttiglione2010a}. Our strong AGNs can be roughly identified with HERGs. Our weak AGNs would belong to the LERG class, as well as the liny and lineless retired galaxies that MIRAD or DLM classify as radio AGNs. It is to be noted that for the majority of the LERGs it is thus not possible to measure the bolometric luminosity of the AGN using optical emission lines. This obviously concerns the retired galaxies, but even in weak AGNs the emission lines are strongly contaminated by emission due to ionization by hot low-mass evolved stars (HOLMES) and not by an AGN \citep[see][]{Fernandes2011}.

\subsection{The relation with other citeria to identify radio AGNs}
\label{otherclas}

In addition to DLM, several other diagrams have been used in the past to help identify radio-AGNs in radio surveys. These are the BPT diagram, the \LHa versus \Lrad\ diagram \citep{Best.Heckman.2012a, Heckman.Best.2014a} and the WISE colour-colour diagram used by \citet{Sabater19}. 
In Fig.~\ref{fig:classif} we show these diagrams colour-coding the points according to their classification in the MIRAD diagram for the \rw sample, (panels a--c) and to the DLM diagram for the ROGUE~I sample (panels d--f). For a and d we required  $S/N > 2$ in all emission lines in the BPT, for panels b and e, we required $S/N > 2$ for H$\alpha$, for panel c and f we required  $S/N \geqslant 2$ in all the WISE bands. 

As obvious from the previous sections, classifications using emission lines will miss a large number of radio-AGNs. On the other hand, adding information from emission lines to the one obtained from the MIRAD or DLM diagrams will not improve the classification, since the latter two are clearcut while both the BPT diagrams and the \LHa versus \Lrad\ diagram show a fuzzy separation between radio-AGNs and other objects, as seen in  Fig.   \ref{fig:classif}.
As has been shown by \citet{Stasinska2008,Stasinska.etal.2015a}, the BPT diagram, in addition to requiring good $S/N$ in at least four emission lines,  cannot distinguish galaxies with low-excitation active nuclei from retired galaxies (RGs), i.e.\ galaxies that have stopped forming stars and are ionized by the HOLMES from their old stellar populations. The WISE colour--colour diagram was used by \citet{Sabater19} to  classify the sources according to a simple division at a given value of $W2-W3$ first proposed by \citet{Herpich2016}. \citet{Sabater19} considered this diagnostic rather crude and used it mainly when their other diagnostics gave intermediate or contradictory classifications.


\section{Summary}
\label{summary}

Considering the ROGUE I catalog, a sample of over 32\,000 radio sources showing a radio core associated with optical galaxies, we have provided a simple way to separate radio AGNs from galaxies where the radio emission is a result of recent star formation. This technique will also be applicable to ROGUE II,  the upcoming supplement to the ROGUE~I, consisting of the SDSS radio galaxies without a close ($<$3\arcsec) FIRST detection. It can also be applied to other catalogs of radio sources, such as LoTSS. 

While previous works \citep{Best.Heckman.2012a,Sabater19} used a conjunction of criteria to sieve out radio-AGNs in catalogs of radio sources, we plead for the use of just one criterion. This provides a simpler diagnostic and a better evaluation of selection effects.

We propose two different diagnostic diagrams which are equally efficient and can be used on large samples of radio sources. 

One is  $\log F_{W3}$ vs $\log \Frad$, dubbed the MIRAD diagram. This diagram only requires radio fluxes and photometry in the WISE $W3$ band with S/N $\geqslant$ 2. It does not require any redshift determination. The MIRAD diagram  neatly separates the radio sources 
in two branches. This was already found by \citet{Rosario13} on a sample of optical AGNs and \citet{Mingo2016} on a sample of X-ray AGNs. Here we showed that the separation subsists when including purely star-forming galaxies  and objects that cannot be optically classified as AGNs due to the lack of emission lines, making it extremely useful to extract radio AGNs from radio catalogs in which many sources are unresolved. 
The other diagram is  $D_{\rm n}(4000)$ versus $\Lrad/\Mstar$, which we call the DLM diagram. This diagram has already been used in previous studies, but always in conjunction with other criteria. Here, we argue that it does a perfectly good job when used alone. 

For these two diagrams, we propose simple, empirical dividing lines that provide the same classification  by both diagrams for the objects in common. These dividing lines classify correctly 99.5 percent of extended  radio sources in the ROGUE~I catalog into the radio-AGN class, and 98--99 percent of the BPT star forming galaxies (using the dividing line of \citet{Stasinska.etal.2006a}) into the SF class. In the ROGUE~I catalog, the DLM diagram can be applied to a larger number of radio sources than the MIRAD diagram because it makes no use of mid-infrared data. On the other hand, the DLM diagram requires optical spectra, which are not necessary available for galaxies in radio catalogs (e.g. the LoTSS catalog). When it can be used, one must recall, however, that it may be affected by aperture effects. For example, in the case of observation of a nearby object ($z \lesssim 0.05$) the light  collected by the SDSS fiber is dominated by the emission from a `retired bulge' \citep[see][]{Herpich2016,Gomes2016a}.

Both the MIRAD and the DLM diagrams clearly illustrate the well known observational fact that radio AGNs are preferentially found among galaxies with the most massive black holes. The distribution of the objects with different optical morphologies in the MIRAD and DLM diagram show that most radio-AGNs correspond to elliptical galaxies.

About 90 per cent of the radio sources classified in MIRAD or DLM diagrams as radio-AGNs are optically weak AGNs or retired galaxies, ie. galaxies that have stopped forming stars and are ionized by photons arising from their hot low-mass evolved stars. In the latter case, if the \Ha\ line is detected, it is not related to the activity of the black hole. 

\section*{ACKNOWLEDGMENTS}

We thank \L{}ukasz Stawarz for discussions. This work was started within the framework of the Polish National Science Centre (NCN) grant 2013/09/B/ST9/00026. DKW and MS acknowledge the support of NCN grant via 2016/21/B/ST9/01620. MS acknowledges the partial support of NCN grant via 2015/18/A/ST9/00746. NVA acknowledges support of the Royal Society and the Newton Fund via the award of a Royal Society--Newton Advanced Fellowship (grant NAF\textbackslash{}R1\textbackslash{}180403), and of  Funda\c{c}\~ao de Amparo \`a Pesquisa e Inova\c{c}\~ao de Santa Catarina (FAPESC) and Conselho Nacional de Desenvolvimento Cient\'{i}fico e Tecnol\'{o}gico (CNPq). GS and NVA acknowledge financial support from the Jagiellonian University, GS acknowledges further support from Paris Observatory. FRH thanks FAPESP for the financial support through the project 2018/21661-9. The work of N.\.{Z}. is supported by the South African Research Chairs Initiative (grant no. 64789) of the Department of Science and Innovation and the National Research Foundation\footnote{Any opinion, finding and conclusion or recommendation expressed in this material is that of the authors and the NRF does not accept any liability in this regard.} of South Africa. AG acknowledges NCN through the grant 2018/29/B/ST9/02298.

    Funding for the SDSS and SDSS-II has been provided by the Alfred P. Sloan Foundation, the Participating Institutions, the National Science Foundation, the U.S. Department of Energy, the National Aeronautics and Space Administration, the Japanese Monbukagakusho, the Max Planck Society, and the Higher Education Funding Council for England. The SDSS Web Site is http://www.sdss.org/.
    The SDSS is managed by the Astrophysical Research Consortium for the Participating Institutions. The Participating Institutions are the American Museum of Natural History, Astrophysical Institute Potsdam, University of Basel, University of Cambridge, Case Western Reserve University, University of Chicago, Drexel University, Fermilab, the Institute for Advanced Study, the Japan Participation Group, Johns Hopkins University, the Joint Institute for Nuclear Astrophysics, the Kavli Institute for Particle Astrophysics and Cosmology, the Korean Scientist Group, the Chinese Academy of Sciences (LAMOST), Los Alamos National Laboratory, the Max-Planck-Institute for Astronomy (MPIA), the Max-Planck-Institute for Astrophysics (MPA), New Mexico State University, Ohio State University, University of Pittsburgh, University of Portsmouth, Princeton University, the United States Naval Observatory, and the University of Washington.

    This publication makes use of data products from the Wide-field Infrared Survey Explorer, which is a joint project of the University of California, Los Angeles, and the Jet Propulsion Laboratory/California Institute of Technology, funded by the National Aeronautics and Space Administration.

    This research made use of Astropy,\footnote{Astropy Python package: \url{http://www.astropy.org}} a community-developed core Python package for Astronomy \citep{AstropyCollaboration.etal.2013a, AstropyCollaboration.etal.2018a}.

\bibliography{references}{}
\bibliographystyle{aasjournal}

\label{lastpage}

\end{document}